\documentclass[12pt,a4paper]{article}

\usepackage{amssymb}
\usepackage{amsmath}
 
\newcommand{\be}{\begin{equation}}
\newcommand{\ee}{\end{equation}}
\newcommand{\bea}{\begin{eqnarray}}
\newcommand{\eea}{\end{eqnarray}}
\newcommand{\lb}{\label}
\newcommand{\bdm}{\begin{displaymath}}
\newcommand{\edm}{\end{displaymath}}
\newcommand{\D}{{\rm d}}

\begin{document}

\begin{titlepage}

\noindent
\begin{center}
\vspace*{1cm}

{\large\bf ON THE CONCEPT OF LAW IN PHYSICS}\footnote{To appear in the
  Proceedings of the conference {\em The Concept of Law in Science},
  Heidelberg, 4--5 June 2012.}  
 
\vskip 1cm

{\bf Claus Kiefer} 
\vskip 0.4cm
Institute for Theoretical Physics,\\ University of Cologne, \\
Z\"ulpicher Strasse~77,
50937 K\"oln, Germany.\\ {\tt http://www.thp.uni-koeln.de/gravitation/}
\vspace{1cm}

\begin{abstract}
I discuss the main features of the concept of law in physics. 
Fundamental laws from Newtonian mechanics to general relativity are
reviewed. I end with an outlook on the new form of laws in the
emerging theory of quantum gravity. 
\end{abstract}

\end{center}

\end{titlepage}


\section{Laws of Nature}

The concept of law is widespread in both the sciences and the
humanities. When one talks about laws of Nature, however,
one usually refers to physics. What is a physical law? Richard
Feynman, in his well known book {\em The Character of Physical Law}
writes (\cite{Feynman}, p.~13):
``There is also a rhythm and a pattern between the phenomena of nature
which is not apparent to the eye, but only to the eye of analysis; and
it is these rhythms and pattern which we shall call Physical Laws.''
As a prototype of a physical law, Feynman states the law of gravitation. 

As is evident from Feynman's quote, one needs a certain degree of
abstraction from the phenomena to discern the laws of Nature. Without
the eye of analysis, physical laws cannot be found. 

In this essay, I shall briefly summarize the status of laws of Nature
in modern physics and speculate about the development of new laws. A
central role is there indeed played by gravitation. On the one hand,
Einstein's theory of general relativity has introduced a dynamical
spacetime into physics and has thus dramatically changed our attitude
towards the formulation of fundamental laws. On the other hand, one
expects that the consistent unification of general relativity with
quantum theory will lead to a completely new type of law. For this
reason, I shall discuss  some aspects of fundamental laws 
as they appear in one approach to quantum gravity.

Physical laws are formulated within a physical theory or a set of
theories. A theory consists of a set of mathematical equations and a
set of mapping rules to phenomena in Nature. In the ideal case, these
rules include a statement about their domain of validity. 

In his famous book {\em Il Saggiatore}, Galileo Galilei has introduced
the picture of Nature as a book written in mathematical
language. This must, however, not be interpreted too literally. The
mathematical language is not unique, and the same phenomena can be
described in different ways. A good example is gravitation.
 In Newtonian terms, the motion of
planets is described by differential equations containing an action at
a distance. In Einsteinian terms, the planets move on geodesics in
spacetime. If gravity is combined with quantum theory, yet another
mathematical picture emerges. There is thus not a one-to-one
relation between mathematics and reality. This is clearly expressed by
a famous quote from Albert Einstein, who writes (\cite{Einstein},
p.~119--120)   

\begin{quote}
Insofern sich die S\"atze der Mathematik auf die Wirklichkeit beziehen,
sind sie nicht sicher, und insofern sie sicher sind, beziehen sie sich
nicht auf die Wirklichkeit.\footnote{In so far the theorems of
  mathematics refer to reality, they are not certain, and in so far
  they are certain, they do not refer to reality.}
\end{quote}

According to Einstein, a certain degree of intuition is needed to find
the correct laws of Nature; they cannot just be read off from the
phenomena. Still, physical laws are not invented, but discovered,
because they reflect properties of the real world, not just our
imagination. In contrast to this, mathematical concepts are, in my
opinion, invented. Why there are laws of Nature at all, is not
obvious; nor is it a priori clear that we are able to discover them.

One can distinguish between physical laws at different levels. Here,
we are mainly concerned with the fundamental laws, that is, laws that
describe the fundamental interactions; examples are the laws of
gravitation and electrodynamics. It is an open issue whether all these
fundamental laws can be unified to one fundamental theory, often
called `theory of everything'. If this happened, it would be the
ultimate triumph of the reductionist programme in physics. 

At a different level, one has effective laws such as the Second Law of
thermodynamics. As we shall briefly discuss below, the Second Law
seems to be a consequence of particular boundary conditions of our
world, and it is open whether it can be derived in a different way
from structures of a new theory, such as quantum gravity.
  
Yet another level concerns emergent laws for complex systems.
They can, in principle, be derived from the fundamental laws, but show
features that go much beyond those laws. 
In the words of Paul Anderson (\cite{Anderson}, p.~395), 

\begin{quote}
\ldots the whole becomes not only more than but very different from
the sum of its parts.
\end{quote}

Here, we shall not discuss such emergent laws, but focus on the
fundamental physical laws from Newtonian mechanics to quantum gravity.


\section{From Newtonian Mechanics to Special Relativity}

A most important feature of our physical theories is the separation of
the description into dynamical laws and initial conditions. This was
expressed very clearly in Eugene Wigner's Nobel speech (\cite{Wigner},
p.~7--8),

\begin{quote}

The regularities in the phenomena which physical science endeavors to uncover
are called the laws of nature. \ldots
The elements of the behavior
which are not specified by the laws of nature are called initial
conditions. \ldots
The surprising discovery of Newton's age is just the clear
separation of laws of nature on the one hand and initial conditions on
the other. The former are 
precise beyond anything reasonable; we know virtually nothing about the
latter.

\end{quote}

Mathematically, our fundamental laws can be expressed as differential
equations up to second order in space and time. They leave thus room
for initial or (more generally) boundary conditions. 
Alternatively, the same laws can be expressed in integral form, as a
variational principle, but this form is fully equivalent to the
differential form. 

Because the physical laws can be formulated as differential equations, they are 
completely deterministic. Determinism, in this modern sense, must not
be confused with causality. If temporal boundary conditions are
specified at a particular time, the solution of the equations is
determined for any time, both before and after that particular time. 
Determinism is one of the most important concepts when discussing
physical laws \cite{Huettemann,MV00}. 

Determinism does not yet mean predictability. Predictability
presupposes determinism, but not the other way around
\cite{Huettemann}. Most systems in Nature exhibit chaotic behaviour,
which means that small perturbations become exponentially
large. Because this is not in conflict with determinism, one talks
about `deterministic chaos'. The prediction of the weather is a
classic example, but already systems as simple as a double pendulum
show chaotic behaviour. For the same reason, it is also not possible to
predict the future of our Solar system, that is, the future of the
motion of planets and asteroids, for more than about four million
years. 

Fundamental physical laws refer to dependences on space and time. It
was one of Newton's great achievement to introduce the concepts of absolute
space and absolute time to facilitate the formulation of his laws. To
quote from his {\em Principia} (\cite{Barbour1}, p.~623),

\begin{quote}
Absolute, true, and mathematical time, of itself, and
from its own nature, flows
equably without relation to anything external. \ldots
Absolute space, in its own nature, without relation to anything
external, remains always similar and immovable. \ldots
\end{quote}

Let us consider Newton's second law of motion for the motion of a set
of $N$ particles described by their positions ${\mathbf x}_i$,
$i=1,\ldots,N$, 
\be
\lb{Newton}
m_i\frac{\D^2{\mathbf x}_i}{\D t^2}= {\mathbf F}_i.
\ee
The force ${\mathbf F}_i$ on the $i$th particle is here assumed to be {\em
  given}. In the 
important case of gravitational interaction, it reads

\be
{\mathbf F}_i=-G\sum_{j\neq i}\frac{m_im_j}{\vert{\mathbf
    x}_i-{\mathbf x}_j\vert^2}\frac{{\mathbf x}_i-{\mathbf
    x}_j}{\vert{\mathbf x}_i-{\mathbf x}_j\vert}. 
\ee

Because \eqref{Newton} is a differential equation of second order in
time, its solution is determined if position and velocity are
specified at a particular moment of time. 

The notions of absolute space and absolute time were criticized at
several occasions in the history of science, mainly because these
notions involve absolute (non-dynamical) elements. Among the critics
were Berkeley, Leibniz, and Mach. Since, however, 
Newton's mechanics was extremely successful, attempts to formulate an
alternative mechanics did not go very far \cite{Barbour1}. Only after
the advent of general relativity did people investigate models of
classical mechanics without absolute space and time \cite{Barbour2}.

Besides gravitation, the only fundamental interaction that manifests
itself at a macroscopic level is electrodynamics. It is described by
the set of Maxwell's equations,
\begin{eqnarray}
\nabla{\mathbf B}=0\ \;\;\;&,& \quad \nabla\times{\mathbf E}+
                       \frac{1}{c}
        \frac{\partial{\mathbf B}}{\partial t}=0\ ,\nonumber \\
\nabla{\mathbf E}=4\pi\rho\ &,& \quad \nabla\times{\mathbf B}
-\frac{1}{c}\frac{\partial{\mathbf E}}{\partial t}
=\frac{4\pi}{c}{\mathbf j},
\end{eqnarray}
where ${\mathbf B}$ and ${\mathbf E}$ are the magnetic and electric
field, respectively. In contrast to Newton's equations, these are
equations for local fields. Already Maxwell's contemporaries were
impressed by the fact that these equations encode all the phenomena
related to electricity, magnetism, and optics.\footnote{It was
  Boltzmann who cited Goethe's {\em Faust}: ``War es ein Gott, der
  diese Zeichen schrieb?''.} One of the main features of new
fundamental laws is the fact that they can predict the occurrence of
new phenomena. In the case of Maxwell's equations, these include the
generation of radio waves, which proved to be of enormous
technological significance.  

In the formulation of physical laws, symmetry principles play a key
role. Otherwise, it would almost be impossible to devise the correct
equations out of the immense number of mathematical options. In
classical mechanics, an important principle is the principle of
relativity: the physical laws are invariant with respect to the
transformation from one inertial frame into another. Maxwell's
equations seem to violate this principle, because they contain a
distinguished speed -- the speed of light $c$. It was this apparent
conflict between mechanics and electrodynamics that led Albert
Einstein in 1905 to his special theory of relativity. By a careful
analysis of the concept of time, he realized that Maxwell's equations
do indeed obey the relativity principle, although the transformation
law becomes more complicated (Lorentz instead of Galileo
transformations). As Hermann Minkowski found out in 1908, special
relativity can most clearly be formulated in terms of a
four-dimensional union of space and time called spacetime
(later simply called Minkowski space).
Time by itself and space by itself are no longer absolute, but
spacetime still is; its main characteristic is the presence of the lightcone
structure, because the speed of light is the same in all inertial
frames. 
 It is then mandatory to formulate all physical laws in
 four-dimensional form so that covariance with respect to
 transformations between inertial systems is evident. Otherwise, it is
 not clear if new laws satisfy the relativity principle or not.   

Quantum mechanics has introduced many new concepts into physics, but
with respect to time nothing has changed; the theory 
has inherited Newton's absolute time.
The time parameter $t$ that occurs in the Schr\"odinger equation,
\be
\lb{Schroedinger}
\hat{H}\Psi={\rm i}\hbar\frac{\partial\Psi}{\partial t},
\ee
is nothing but the time parameter of \eqref{Newton}. 

The Schr\"odinger equation \eqref{Schroedinger} is a deterministic
equation: if the quantum state $\Psi$ is given at any particular 
instant of time,
the solution follows for any other time value, both before and after
that instant. The interpretation of $\Psi$ is, however, drastically
different from classical fields such as ${\mathbf E}$ or ${\mathbf
  B}$, because it is defined not in spacetime, but on a
high-dimensional configuration space. Its connection with classical
quantities is described by the probability interpretation. The
emergence of classical behaviour is given by the process of
decoherence \cite{deco}.

If special
relativity is combined with quantum theory, one arrives at quantum
field theory. Here, four-dimensional flat Minkowski space is
used as a rigid classical background on which the dynamics of the
quantum fields is defined. 

As many authors, in particular Albert Einstein, have noted, it is not
natural to envisage something that can act but which cannot be acted
upon (as is the case for Minkowski space). 
The situation changes dramatically with general relativity, to which
we now turn.


\section{General Relativity}

In general relativity, the gravitational field is described by the
geometry of a dynamical four-dimensional spacetime. The fundamental
equations are a set of ten coupled partial differential equations for
the metric $g_{\mu\nu}$. In standard notation, these Einstein field equations
read
\begin{equation}
\lb{Einstein}
R_{\mu\nu}-\frac{1}{2}g_{\mu\nu}R+\Lambda g_{\mu\nu}
= \frac{8\pi G}{c^4}T_{\mu\nu}.
\end{equation}
For the first time, one is confronted with equations that are not
formulated on a given spacetime, but equations that describe spacetime
itself. One impressive example is the existence of gravitational waves,
which describe the propagation of pure curvature without matter. 
As John Wheeler always emphasized, space tells matter how to move, and
matter tells space how to curve.

In spite of the complex nature of the Einstein field equations, a
well-defined initial value problem (`Cauchy problem') can be
formulated. The metric coefficients $g_{\mu\nu}(x)$ can be determined
uniquely (up to coordinate transformations) from appropriate initial
data. An important feature in this context is the presence of
four (at each space point) {\em constraints}. These constraints
 arise from the fact that the theory is invariant
under four-dimensional diffeomorphisms (`coordinate
transformations'). The initial data consist of the three-dimensional
metric, the second fundamental form, and matter degrees of freedom on
a spacelike hypersurface that satisfy the four constraints. 
In this way, spacetime itself is constructed from initial data.
The existence of a well-defined Cauchy problem is of special relevance
for numerical relativity, which is concerned with processes such as the
evolution of two black holes orbiting each other. 

General relativity is a very successful theory. With perhaps the
exception of dark matter and dark energy, it describes all known
gravitational phenomena. But it behaves also in an exemplary manner 
with respect to its limits. From general theorems (`singularity
theorems'), one knows that there are situations in which the theory
breaks down \cite{HP96}. These are, in fact, important situations
because they apply to the origin of our Universe and to the interior
of black holes. For these and other reasons, one expects that the laws
of gravity are, at the most fundamental level, not exactly described by 
Einstein's equations. One way to arrive at a more fundamental theory
than general relativity is to take quantum theory into account. This
will be the subject of the last section.


\section{Quantum Gravity and Beyond}

General relativity and quantum theory cannot both be exactly valid.
One reason is what usually is referred to as the `problem of time'
\cite{OUP}. 
Time is absolute in quantum mechanics (spacetime in quantum field
theory), but it is dynamical in general relativity (as part of the
dynamical spacetime). So what happens in situations where both
theories become relevant?

If one keeps the linear structure of quantum theory and looks for a
quantum wave equation that gives back Einstein's equations in the
semiclassical limit, one arrives at a quantum constraint equation of
the general form
\be 
\lb{WDW}
\hat{H}\Psi=0.
\ee
This equation is known as the Wheeler--DeWitt equation \cite{OUP}.
It has some amazing properties. The full quantum state $\Psi$ of
gravity and matter depends on the three-dimensional metric only, but
is invariant under three-dimensional coordinate transformations. It
does not contain any external time parameter $t$. The reason for this
`timeless' nature is obvious. In general relativity, a
four-dimensional spacetime is the analogue to a particle trajectory in
mechanics. After quantization, the trajectory vanishes, and so does
spacetime. What remains is space, and the configuration space is the
space of all three-geometries \cite{essay}. A constraint equation of
the form \eqref{WDW} also occurs in loop quantum gravity \cite{OUP}. 

To give a particular example, let us formulate the Wheeler--DeWitt
equation for a simple cosmological model. For a closed
Friedmann--Lema\^{\i}tre universe with scale factor $a$ and a massive
scalar field $\phi$, this equation reads after an appropriate choice
of units as follows ($\Lambda$ is the cosmological constant):
\be
\frac{1}{2}\left(\frac{\hbar^2}{a^2}\frac{\partial}{\partial a}
\left(a\frac{\partial}{\partial a}\right)-\frac{\hbar^2}{a^3}
\frac{\partial^2}{\partial\phi^2}-a+\frac{\Lambda a^3}{3}+
m^2a^3\phi^2\right)\psi(a,\phi)=0.
\ee
The timeless nature is evident. The cosmological wave function
only depends on the two variable $a$ and $\phi$. As can be seen from
the kinetic term, the Wheeler--DeWitt equation is of hyperbolic nature
(this is also true for the general case). It does provides the means to
define an {\em intrinsic time}, which is distinguished by the sign in
the kinetic term. This intrinsic time, however, is no longer a time
given from the outset, but is defined from the three-dimensional
geometry itself. In this way, it resembles what the astronomers used
to call ephemeris time \cite{Barbour2}.  

The Wheeler--DeWitt equation thus represents a new type of physical
law. It describes a timeless world at the most fundamental level. The
usual time parameter of physics emerges only at an approximate level
and under very special circumstances \cite{OUP}. 
To quote John Wheeler from his pioneering work (\cite{Wheeler}, p.~253),

\begin{quote}
These considerations reveal that the concepts of spacetime and time
itself are not primary but secondary ideas in the structure of
physical theory. These concepts are valid in the classical
approximation. However, they have neither meaning nor application
under circumstances when quantum-geometrodynamical effects become
important. \ldots There is no spacetime, there is no time, there is no
before, there is no after. The question what happens ``next'' is
without meaning. 
\end{quote}

In spite of its timeless nature, the Wheeler--DeWitt equation can, in
principle, provide the means for an understanding of the arrow of time
(the Second Law of thermodynamics mentioned above) \cite{Zeh}. 
Let us consider a
Friedmann--Lema\^{\i}tre universe with scale factor $a\equiv
\exp(\alpha)$ and small perturbations symbolically denoted by
$x_n$. The Wheeler--DeWitt equation then assumes the form
\begin{equation}
\hat{H}\Psi=\left(\frac{\partial^2}{\partial\alpha^2}
+\sum_n\left[-\frac{\partial^2}{\partial x_n^2}+
\underbrace{V_n(\alpha,x_n)}_{\to 0 \ 
\mbox{for}\ \alpha\to -\infty}\right]\right)\Psi=0.
\end{equation}
The potentials $V_n$ have the property that they vanish when the scale
factor goes to zero (i.e., near the big bang and the big crunch). 
This expresses a fundamental asymmetry of the Wheeler--DeWitt
equation. For small scale factor,
therefore, one can have solutions that are fully
unentangled among the degrees of freedom. But for increasing $a$ the
solution becomes entangled, and one can obtain a non-vanishing 
entanglement entropy
upon tracing out the perturbations. Entanglement entropy can be
related to thermodynamic entropy, and this
entropy then increases with
increasing size of the universe and thereby defines a definite
direction. In the limit where an approximate time parameter is
present, this gives rise to the usual Second Law. But if viewed from
this fundamental perspective, the expansion of the universe is a pure
tautology.

So what is the future of physical laws? The Wheeler--DeWitt equation
has not yet been experimentally tested, but it is an equation that
follows in a straightforward way from the unification of quantum
theory with gravity. It describes quantum effects of gravitation, but
does not encompass by itself a unification of all interactions. 
A candidate for a unified theory is string theory. The structure of
the fundamental laws in this approach is not yet fully understood, but
it seems to be different from the structures discussed above \cite{OUP}.
 
Whether a fundamental `theory of everything' can be found, is open. 
It may happen that such a theory will be available in this century and
that the fundamental picture in physics is `complete' in the sense
that all phenomena can be derived from it, at least in principle. Or
it may happen that we are stuck because experimental progress becomes
slower and slower and no decision among candidates for a fundamental
theory can be made. In one way or another, it is true what Feynman
wrote already in 1964 (\cite{Feynman}, p.~172): 
``The age in which we live is the age in which
we are discovering the fundamental laws of nature, and that day will
never come again. It is very exciting, it is marvellous, but this
excitement will have to go.'' 



\end{document}